\documentstyle[12pt,draft,epsf]{article}
\newcommand{\resection}[1]{\setcounter{equation}{0}\section{#1}}

\def\be{\begin{equation}}
\def\ee{\end{equation}}
\def\bea{\begin{eqnarray}}
\def\eea{\end{eqnarray}}
\def\beano{\begin{eqnarray*}}
\def\eeano{\end{eqnarray*}}
\def\bd{\begin{displaystyle}}
\def\ed{\end{displaystyle}}
\def\ba{\begin{array}}
\def\ea{\end{array}}

\begin{document}
\oddsidemargin 5mm
\setcounter{page}{0}
\newpage     
\setcounter{page}{0}
\begin{titlepage}
\begin{center}
{\large {\bf Bosonic Theory with a Random Defect Line}}\\
\vspace{1.5cm}
{\bf M. Moriconi} \footnote{\tt{email:marco@if.ufrj.br}}\\
{\em Instituto de F\'\i sica}\\
{\em Universidade Federal do Rio de Janeiro}\\
{\em Rio de Janeiro, RJ 21945-970, Brazil}\\
\vspace{0.8cm}
\end{center}
\renewcommand{\thefootnote}{\arabic{footnote}}
\vspace{6mm}

\begin{abstract}
\noindent
We study a two-dimensional bosonic field theory with a random defect line. The theory has a background field coupled to the field variables at the defect line, which renders the model non-integrable. However, as the background field is random, and the disorder is implemented through the replica trick, the model becomes integrable, allowing us to use the form-factor method to compute the exact correlation functions of the quenched model. 
\vspace{3cm}

\end{abstract}
\vspace{5mm}
\end{titlepage}

\newpage
\setcounter{footnote}{0}
\renewcommand{\thefootnote}{\arabic{footnote}}

\resection{Introduction} 

Integrable field theories in two-dimensions \cite{zz} are at the same time useful from the theoretical poin of view as well as to applications to statistical mechanics and condensed matter problems \cite{s1,s2}. Recently there has been a considerable amount of work on two-dimensional quantum field theories with boundaries and defect lines \cite{gz}-\cite{jl}. The interest on this type of theories is due to the fact that they can model such systems as quantum impurities, for example. It has been shown that if one wants the theory to  be integrable even after the defect has been introduced, then the bulk theory has, necessarily, a constant $S$-matrix, being basically that of a free fermion or free boson \cite{dms2,cfg}. This restricts quite severely the type of interactions one can use at the defect line in order to mantain integrability.

In this paper we study a two-dimensional free boson with a disordered defect line. Before taking the disorder into account the theory is {\em non-integrable}, but after disordering the line, through the replica trick \cite{b}, we obtain an integrable field theory for all values of $n$, the number of replicas. Another interesting phenomenon that happens concerns the stability of the vacuum of the theory. As it has been shown in \cite{dms2} there is a minimun value for the coupling constant, below which the theory becomes unstable. We will see that this happens for every finite value of $n$, but that in the limit $n \to 0$ the theory becomes well-defined for all values of the coupling constant (which is the width of the disorder $\Delta$).

This paper is outlined as follows. In section 2 we introduce the model to be studied and discuss some of its properties. In section 3 we apply the replica trick and obtain the effective field theory with finite $n$. The reflection-transmission algebra is briefly reviewed in section 4, and in section 5 we compute the reflection and transmission amplitudes.
We show how to compute correlation functions using these amplitudes in the next section, and comment on the analytical structure of the theory as $n \to 0$. Our conclusions are presented in section 7. In the appendix we collect some of the useful formulas to compute correlation functions for the disordered theory in the $n \to 0$ limit.

\section{The Disordered Bosonic Model}

Consider a two-dimensional free-boson Euclidean quantum field theory (qft) with a line of defect, given by the following action
\be
S=\frac{1}{2}\int d^2x \; (\partial \phi)^2 - m^2 \phi^2 + \int dt \; {\cal L}_d(\phi, \partial \phi) \ ,
\ee
where ${\cal L}_d(\phi, \partial \phi)$ is the defect interaction lagrangian. It is clear that, even though the model is obviously solvable in the bulk, we have to choose judiciously the defect interaction in order to mantain integrability. In principle we could have chosen different qft's for $x<0$ and $x>0$, as have Corrigan et al done in classical context \cite{bcz1,bcz2}, but in this case we will consider the same theory on both sides of the defect.

If we couple this theory to an external field with a defect action $\int_{-\infty}^{+\infty} dt \; h(t)\phi(0,t)$, then the theory is clearly non-integrable, since the external field can absorb or produce new particles. We spoil the fact that in an integrable model there is no particle production in a scattering process.

The equations of motion for this model are easily derived to be the usual Klein-Gordon equation, plus a source term
\be
\partial^2 \phi + m^2 \phi = \delta(x) h(t) \ .
\ee
The coupling with the external field is, therefore, taken into account through the boundary condition. In order to consider the (quenched) disordered theory we are going to use the replicated action, as discussed in the next section.

\section{The Disordered Action: Replica Trick}

When considering quenched theories we need to compute the average of the Helmholtz free energy, ${\cal F}$. The technical complication that arises immediately is how to do that, since we need to compute, essentially, the average of the logarithm of the partition function ${\cal Z}$. This may be acomplished through the replica trick: we write the average $\overline{\ln {\cal Z}}=\lim_{n \to 0} \frac{\overline{{\cal Z}^n-1}}{n}$. By taking the average before the limit $n \to 0$, we arrive at the replicated action. Some of the formulas showing how to compute correlation functions from this replicated action are shown in the appendix.

We consider a gaussian distribution for the background field $h$. This implies that $\langle h(t)h(t') \rangle = \Delta \delta(t-t')$ and $\langle h(t) \rangle = 0$. All the higher moments can be calculated from these two expectation values.

The replicated action will be given, then, by
\bea
{\cal Z}_{\rm eff}^{(n)}=\int \prod_{k=1}^n{\cal D} \phi_k
\int {\cal D}h \exp\left(-\sum_{k=1}^n(\frac{1}{2}\int d^2x \; [(\partial \phi_k)^2 - m^2 \phi_k^2]+
\int_{-\infty}^{+\infty} dt \; h(t) \phi_k)-\right. \nonumber \\
\left.-\frac{1}{2\Delta}\int_{-\infty}^{+\infty} dt \; h(t)^2\right) \ .
\eea
Performing the integration over the background field, we obtain
\be
{\cal Z}_{\rm eff}^{(n)}=\int \prod_{k=1}^n{\cal D} \phi_k \exp\left( (-\frac{1}{2}\sum_{k=1}^n(\int d^2x \; [(\partial \phi_k)^2 - m^2 \phi_k^2])+\frac{\Delta}{2}\int_{-\infty}^{+\infty} dt \; (\sum_{k=1}^n \phi^{(k)})^2\right) \ . \label{zeff}
\ee
This partition function defines the action we are going to study in the next sections. We can readily see that the model has become integrable: the defect interaction becomes quadratic in the replica copies. An analogous mechanism takes place in the case of the ising model with a random boundary magnetic field \cite{dmm}.

\section{Reflection-Transmission Algebra}

Similarly to the case of two-dimensional boundary integrable field theories \cite{gz}, we can start from a given $S$-matrix, and look for the constraints imposed on the reflection and transmission amplitudes. In what follows we will consider diagonal $S$-matrices, for simplicity. We also assume that the defect has no internal degrees of freedom, the inclusion of which is rather elementary. Let us denote the reflection and transmission amplitudes by $R_{ij}(\theta)$ and $T_{ij}(\theta)$, and let $Z_i(\theta)$ be the Faddeev-Zamolodchikov operator that creates an asymptotic state for particle $i$ with rapidity $\theta$.
\vskip 0.5cm
\centerline{\epsffile{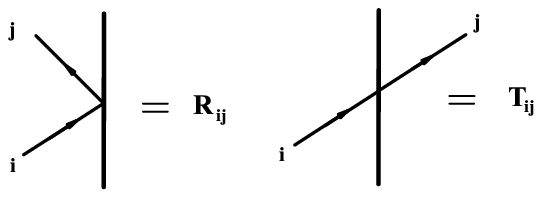}}
\vskip 0.5cm
\centerline{Fig 1. Graphical representation of reflection and transmission amplitudes.}
\vskip 0.5cm
Since the model we are considering does not break parity, we do not have to make a distinction between amplitudes of particles reflecting or scattering from the right or left side of the defect.
Let us introduce the operator ${\cal D}$ for the defect. The operators $Z_i(\theta)$
and ${\cal D}$ obey the following algebra
\bea
&&Z_i(\theta){\cal D}=R_{ij}(\theta)Z_j(-\theta){\cal D}+T_{ij}(\theta){\cal D}Z_j(\theta) \ , \nonumber \\
&&{\cal D}Z_i(\theta)=R_{ij}(-\theta){\cal D}Z_j(-\theta)+T_{ij}(-\theta)Z_j(\theta){\cal D} \ .
\eea
Applying this algebra twice in ${\cal D} Z_i(\theta)$, implies the following unitary conditions
\bea
&&R_{ik}(\theta)R_{kj}(-\theta)+T_{ik}(\theta)T_{kj}(-\theta)=\delta_i^j \ , \nonumber \\
&&R_{ik}(\theta)T_{kj}(-\theta)+T_{ik}(\theta)R_{kj}(-\theta)=0 \ .
\eea
Since we are considering diagonal $S$-matrices, the Yang-Baxter equation is trivially satisfied.

\section{Reflection and Transmission Amplitudes}

Once we have the equations of motion of the disordered model (\ref{zeff}) we can use a mode decomposition of the left and right fields in order to solve them and find the reflection and transmission amplitudes.

The $k$-th copy of the field in the replicated action, $\phi^{(k)}$, can be written as a sum of fields to the left and to the right of the defect
\be
\phi^{(k)}(x,t)=\theta(x)\phi_+^{(k)}(x,t)+\theta(-x)\phi_-^{(k)}(x,t) \ , \label{phi+-}
\ee
where $\theta(x)$ is the Heaviside function.

Let us introduce the mode decomposition for each replica as follows
\be
\phi_{\pm}^{(k)}(x,t)=\int_{-\infty}^{+\infty} \frac{d\theta}{2\pi}\left( a_\pm^{(k)}(\theta)\exp(-im(t\cosh\theta - x\sinh\theta))+ {\rm c.c.}\right) \ , \label{modes}
\ee
where $\theta$ is the rapidity variable, meaning that energy and momentum are written as $(e,p)=(m\cosh \theta, m\sinh\theta)$, and the coefficients $a_\pm^{(k)\dagger}(\theta)$ and $a_\pm^{(k)}(\theta)$ are the creation and annihilation operators, respectively, and they satisfy the usual commutation relations
\be
[a_\pm^{(i)}(\theta_1),a_\pm^{(j)\dagger}(\theta_2)]=2\pi\delta^{ij}\delta(\theta_1-\theta_2)
\ , \ee
and all the remaining commutators vanish.

Substituting the decomposition (\ref{phi+-}) into the equation of motion, we obtain the following boundary conditions, by integrating around $x=0$, and using appropriate regularizations for the Heaviside and delta functions
\bea
&&\partial_x(\phi_+^{(k)}(0,t)-\phi_-^{(k)}(0,t))=
\frac{\Delta}{4}\sum_{j=1}^{n}(\phi_+^{(j)}(0,t)+\phi_-^{(j)}(0,t)) \ , \nonumber \\
&&\phi_+^{(k)}(0,t)=\phi_-{(k)}(0,t) \ .
\eea
We see then that the different replica copies interact through the defect only. These boundary conditions among the field components become conditions among the modes in (\ref{modes})
\bea
&&(a_+^{(k)\dagger}(\theta)-a_+^{(k)\dagger}(-\theta)-a_-^{(k)\dagger}(-\theta)+a_-^{(k)\dagger}(-\theta))=
-\frac{i\Delta}{4m\sinh(\theta)}\sum_{j=1}^{n}(a_+^{(j)\dagger}(\theta)+a_+^{(j)\dagger}(-\theta)+
\nonumber \\
&&\phantom{(a_+^{(k)\dagger}(\theta)-a_+^{(k)\dagger}(-\theta)-a_-^{(k)\dagger}(-\theta)+a_-^{(k)\dagger}(-\theta))=-\frac{i\Delta}{4m\sinh\theta}\sum_{j=1}^{n}(}
a_-^{(j)\dagger}(\theta)+a_-^{(j)\dagger}(-\theta))
\ , \nonumber \\
&&a_+^{(k)\dagger}(\theta)+a_+^{(k)\dagger}(-\theta)=a_-^{(k)\dagger}(\theta)+
a_-^{(k)\dagger}(-\theta) \ ,
\eea
and all the remaining commutators vanish.
It is easy to see that these equations agree with the ones in \cite{dms2} if we take $n=1$ and make the correspondence $\Delta=g$, where $g$ is the coupling constant of the defect for the bosonic theory studied in their paper.

These boundary conditions can be summarised into one matrix equation as follows
\be
M_+[\theta]V[\theta]=M_-[\theta]V[-\theta] \label{MVMV} \ ,
\ee
where $V[\theta]^{\rm T}=[\matrix{a_-^{(1)\dagger} (\theta) & a_+^{(1)\dagger} (-\theta) & \ldots & a_-^{(n)\dagger} (\theta) & a_+^{(n)\dagger}(\theta)}]$ is a $2n$ component vector, and the matrix $M_\pm[\theta]$ is given by
\be
M_\pm[\theta]=
\left[\matrix{\pm 1 & \mp 1 & 0 & 0 & 0 & \ldots \cr 0 & 0 & \pm 1 & \mp 1 & 0 & \ldots \cr
\vdots & \vdots &  &  &  \cr
1\pm h(\theta) & 1 \pm h(\theta) & \pm h(\theta) & \pm h(\theta) & \pm h(\theta) & \ldots \cr
\pm h(\theta) & \pm h(\theta) & 1 \pm h(\theta) & 1 \pm h(\theta)& \pm h(\theta) & \ldots \cr
\vdots & \vdots &  &  &  \cr}\right] \ , \label{M}
\ee
where the first $n$ rows are made out of $\pm 1$'s and the last $n$ rows of $1\pm h$ and $\pm h$, and $h(\theta)=i\Delta/4m\sinh\theta$.
We can find the reflection and transmission amplitudes by using the algebraic method. First, recall that we have a linear relationship among the creation and annihilation operators that act on the right or left of the defect, as follows  
\be
\left[\matrix{a_-^{(1)\dagger} (\theta) \cr a_+^{(1)\dagger} (-\theta) \cr \vdots \cr
a_-^{(n)\dagger} (\theta) \cr a_+^{(n)\dagger} (-\theta)}\right]=
\left[\matrix{R_1^1(\theta) & T_1^1(\theta)& \ldots & R_n^1(\theta) & T_n^1(\theta)  \cr
T_1^1(\theta) & R_1^1(\theta) & \ldots & T_n^2(\theta) & R_n^2(\theta) \cr
\vdots & \vdots & & \vdots & \vdots \cr R_n^1(\theta) & T_n^1(\theta) &\ldots & R_n^2(\theta) & T_n^2(\theta) \cr
T_n^1(\theta) & R_n^1(\theta) & \ldots & T_n^2(\theta) & R_n^2(\theta)
}\right]
\left[\matrix{a_-^{(1)\dagger} (-\theta) \cr a_+^{(1)\dagger} (\theta) \cr \vdots \cr
a_-^{(n)\dagger} (-\theta) \cr a_+^{(n)\dagger} (\theta)}\right] \ .
\ee
The physical interpretation of the $R_i^j(\theta)$ and $T_i^j(\theta)$ is quite simple. The $R_i^j(\theta)$ are the amplitude of probability that an incoming particle of type $i$ moving to the right (left) with rapidity $\theta$ ($-\theta$)reflects to the left (right) at the defect and becomes a particle of type $j$. Similarly, the $T_i^j(\theta)$ represent the amplitude of probability that an incoming particle of type $i$ moving to the right (left) with rapidity $\theta$ ($-\theta$)is transmitted to the left (right) of the defect and becomes a particle of type $j$.

Using equations (\ref{MVMV}) we see that the reflection and transmission amplitudes are obtained from the matrix $M_\pm[\theta]$ by computing $R[\theta]=M_+^{-1}[\theta]M_-[\theta]$.

Using (\ref{M}) we can solve for the $R_{ij}(\theta)$ and $T_{ij}(\theta)$. This is most simply done by noting that the inverse of $M_+^{-1}(\theta)$ has a structure similar to $M_+^{\rm T}(\theta)$
\be
M_+^{-1}(\theta)=\frac{1}{2}
\left[\matrix{
1 & 0 & 0 & \ldots & a(\theta) & b(\theta) & b(\theta) & \ldots \cr
-1 & 0 & 0 & \ldots & a(\theta) & b(\theta) & b(\theta) & \ldots \cr
0 & 1 & 0 & \ldots & b(\theta) & a(\theta) & b(\theta) & \ldots \cr
0 & -1 & 0 & \ldots & b(\theta) & a(\theta) & b(\theta) & \ldots \cr
\vdots & \vdots & \vdots  & & \vdots & \vdots & \vdots & &}\right] \label{M_+^{-1}} \ ,
\ee
where $a(\theta)$ and $b(\theta)$ are yet to be determined. Requiring that (\ref{M_+^{-1}}) is the inverse of $M_+^{-1}$, leads to the following two equations for $a(\theta)$ and $b(\theta)$
\bea
&&(1+h(\theta))a(\theta)+(n-1)h(\theta)b(\theta)=1 \ , \nonumber \\
&& \;h(\theta)a(\theta)+[(n-1)h(\theta)+1]b(\theta)=0 \ .
\eea
The solution of these simple equations gives
\be
a(\theta)=\frac{1+(n-1)h(\theta)}{1+nh(\theta)} \qquad {\rm and} \qquad b(\theta)=-\frac{h(\theta)}{1+n h(\theta)} \ .
\ee
Now we can easily obtain the reflection and transmission amplitudes. Its final form is easily computed from the formulas above, and we obtain
\be
R(\theta)=
\left[\matrix{A(\theta) & B(\theta)& A(\theta) & \ldots & A(\theta)  \cr
B(\theta) & A(\theta) & A(\theta) & \ldots &  A(\theta) \cr
\vdots & \vdots & &  &  \cr 
A(\theta) & A(\theta) &\ldots & A(\theta) & B(\theta) \cr
A(\theta) & A(\theta) & \ldots & B(\theta) & A(\theta)
}\right] \ , \
\ee
that is, there is a $2 \times 2$ block-diagonal structure, with diagonal elements $A(\theta)$ and off-diagonal elements $B(\theta)$, and {\em all} remaining elements are equal to $A(\theta)$. The functions $A(\theta)$ and $B(\theta)$ are given by
\be
A(\theta)=-\frac{h(\theta)}{1+nh(\theta)} \qquad {\rm and} \qquad B(\theta)=\frac{1+(n-1)h(\theta)}{1+nh(\theta)} \ .
\ee
As one expects, all diagonal elements of $R(\theta)$ are equal, since there is no reason why the reflection of one replica off the defect should be differenct from another. What is somewhat surprising is to find out that some of the other elements are equal. This means, for example, that a particle of type $i$ coming from the left of the defect has the same amplitude of being reflected to the left as a particle of type $i$, as to being transmitted through the defect as any particle different from $i$. We are interested, of course, in 
the limit $n \to 0$, where such considerations loose their meaning, though.

\section{Analytical Structure and Correlation Functions}

Since the amplitudes are $2\pi i$-periodic, we can restrict our attention to the physical strip $-i\pi \leq {\rm Im}(\theta) < i\pi$. Let us look at $A(\theta)$ initially. As far as $0<\Delta<4m/n$, there are two poles on the physical strip, in the imaginary axis. If $\Delta = 4m/n$ these two poles coalesce into one with $\theta = -i\pi2$, and as $\Delta > 4m/n$ there are two poles with imaginary part $-i\pi/2$ and real part different from zero. These correspond to ressonant states, and as $\Delta \to \infty$ these poles move to infinity. But since we are interested in the replica limit of $n \to 0$, we see that there are no ressonances for $0 < \Delta < \infty$.

We could also analyse the case $\Delta < 0$, even though $\Delta$ refers to the width of a gaussian distribution for the defect disorder. We are looking, therefore, at the model obtained in (\ref{zeff}), as the starting point field theory. In this case we have that for 
$-4m/n<\Delta<0$ there are two poles on the imaginary axis. As $\Delta$ decreases and reaches $-4m/n$, these two poles coalesce at $i\pi/2$. Further decreasing of $\Delta$ gives a real part to these two poles, that is, they correspond now to instabilities in the theory. As discussed in \cite{dms2}, one can interpret this instability as the emission of pairs of particles from the defect. But in our case, since eventually $n \to 0$ the bound $-4m/n \to -\infty$, that is, the theory becomes well defined for all values of the ``coupling constant" $\Delta$.

We follow the discussion of \cite{dms2} on how to compute correlation functions. In order to do that we rotate the coordinates, $t \to ix$ and $x \to -it$. Note that since $(e,p)=(m\cosh\theta,\sinh\theta)$, we have to replcae, accordingly, $\theta \to i\pi/2-\theta$. We will denote by a hat the functions of $\theta$ computed at $i\pi/2-\theta$. 
We are interested in computing objects like 
\be
G(x_1,t_1;x_2,t_2,\ldots;x_n,t_n)=\langle T[\Phi(x_1,t_1)\Phi(x_2,t_2)\ldots \Phi(x_n,t_n)]\rangle
\ ,
\ee
where the dynamics of the fields $\Phi(x,t)$ is governed by the Hamiltonian associated with (\ref{zeff}), which includes the defect interaction term, and $T[\ldots]$ denotes the time-ordering operator. We can therefore rewrite the correlation function as
\be
\langle T[\Phi(x_1,t_1)\Phi(x_2,t_2)\ldots \Phi(x_n,t_n)]\rangle = \frac{\langle 0| T[\phi(x_1,t_1)\phi(x_2,t_2) \ldots {\cal D} \ldots \phi(x_n,t_n)]|0\rangle}{\langle 0|{\cal D}|0\rangle} \ . \label{correlation}
\ee 
On the left hand side of this equation we compute the expectation value of the string of operators using the vacuum of the complete hamiltonian, that is, the bulk part plus the defect interaction. On the right hand side the fields $\phi$ are governed by the bulk hamiltonian only, and the defect interaction is taken into account by introducing the operator ${\cal D}$ at $t=0$. The meaning of this equation using euclidean path integrals (or the transfer matrix in the associated statistical mechanics problem) is quite clear, and corresponds to the insertion of the perturbing operator ${\cal D}=\exp(-\int {\cal L}_d)$ precisely at $t=0$. 

Within this framework, we can compute all correlation functions using the form-factors of the free bulk theory and matrix elements of the defect operator ${\cal D}$, namely $\langle \theta_1,j| {\cal D}|\theta_2,i\rangle$, $\langle 0| {\cal D}|\theta_1,i;\theta_2,j\rangle$, and $\langle \theta_1,i;\theta_2,j| {\cal D}|0\rangle$, where $|\theta,i\rangle$ refers to an asymptotic state with rapidity $\theta$ and type $i$, and so on.
Using the path-integral interpretation of formula (\ref{correlation}), 
we can show that
\bea
\langle \theta_1,j| {\cal D}|\theta_2,i\rangle &=& 2 \pi \delta(\theta_1-\theta_2){\hat T}_{ij}(\theta_1) \nonumber \\
\langle 0| {\cal D}|\theta_1,i;\theta_2,j\rangle &=& 2\pi \delta(\theta_1+\theta_2) {\hat R}_{ij}(\theta_1) \nonumber \\
\langle \theta_1,i;\theta_2,j| {\cal D}|0\rangle &=& 2\pi \delta(\theta_1+\theta_2) {\hat R}_{ij}(\theta_2) \ ,
\eea

We can compute ${\overline{\langle \Phi^2(x,t) \rangle}}$ by using the formulas described in the appendix. In this case we have
\be
{\overline{\langle \Phi^2(x,t) \rangle}}=\lim_{n\to0}\langle \Phi_r^2(x,t) \ , \rangle
\ee
where the index $r$ is chosen at will. In the following we denote the coordinates $(x,t)$ by $z$. Using the method described above, we have
\bea
&&{\overline{\langle \Phi^2(z) \rangle}}=\langle \phi^2(z){\cal D}\rangle\nonumber \\
&&\phantom{{\overline{\langle \Phi^2(z) \rangle}}}=\lim_{n\to0}\sum_k \langle0|\phi_r^2|k\rangle \langle k| {\cal D}|0\rangle \nonumber \\
&&\phantom{{\overline{\langle \Phi^2(z) \rangle}}}=\lim_{n\to0}\frac{1}{2!}\int\frac{d\theta_1}{2\pi}\frac{d\theta_2}{2\pi}\langle0|\phi_r^2|\theta_1,r;\theta_2,r\rangle \langle \theta_1,r;\theta_2,r| {\cal D}|0\rangle \ ,
\eea
where we have inserted a resolution of the identity, in the asymptotic state basis, between $\phi^2_r$ and ${\cal D}$. The matrix element $\langle0|\phi_r^2|\theta_1,r;\theta_2,r\rangle$ can be computed using the mode expansion for the fields
\be
\langle0|\phi_r^2|\theta_1,r;\theta_2,r\rangle=2 e^{-mt(\cosh\theta_1+\cosh\theta_2)+imx(\sinh\theta_1+\sinh\theta_2)} \ , \
\ee
and therefore we obtain
\bea
&&{\overline{\langle \Phi^2(z) \rangle}}=\lim_{n\to0}2\int_{-\infty}^\infty \frac{d\theta}{2\pi}{\hat R}(\theta) e^{-2mt\cosh\theta} \nonumber \\
&&\phantom{{\overline{\langle \Phi^2(z) \rangle}}}=-\frac{\Delta}{2m}\int_{-\infty}^\infty
\frac{d\theta}{2\pi}\frac{e^{-2mt\cosh\theta}}{\cosh\theta} \ .
\eea
The fact that this correlation function does not depend on the coordinate $x$ is easily understood as a consequence of translation symmetry in a direction paralel to the defect.

Using that $\int_0^\infty d\theta e^{-2mt\cosh\theta}=K_0(2mt)$, and that as $u\to\infty$, $K_0(u) \sim \frac{e^{-u}}{\sqrt{u}}$, we find
\be
{\overline{\langle \phi^2(z) \rangle}} \to \frac{e^{-2mt}}{t^{1/2}}
\ee
as $t \to \infty$.

As another interesting example, we can compute the correlation function of operators located at opposite sides of the defect, that is, $\langle \Phi(z_1)\Phi(z_2)\rangle$, where $t_1<0<t_2$. In this case, we need to compute
\bea
&&\overline{\langle \Phi(z_1)\Phi(z_2)\rangle} = \lim_{n\to0}\sum_{i,j}\int \frac{d\theta_1}{2\pi}\frac{d\theta_2}{2\pi} \langle 0 | \phi_r(z_1)|\theta_1,i\rangle
\langle \theta_1,i|{\cal D}|\theta_2,j\rangle \langle \theta_2,j|\phi_r(z_2)|0\rangle \nonumber \\
&&\phantom{\overline{\langle \Phi(z_1)\Phi(z_2)\rangle}}=\lim_{n\to0}
\int \frac{d\theta_1}{2\pi} \langle 0 | \phi_r(z_1)|\theta_1,r\rangle
\langle \theta_1,r|\phi_r(z_2)|0\rangle {\hat T}_{rr}(\theta_1) \nonumber \\
&&\phantom{\overline{\langle \Phi(z_1)\Phi(z_2)\rangle}}=
\int_{-\infty}^{\infty} \frac{d\theta}{2\pi} (1-\frac{\Delta/4m}{\cosh\theta})e^{-2mt\cosh\theta}
\ , \eea
where $t=t_2-t_1$, which is always positive. From this expression 
we can extract the asymptotic behaviour of 
$\langle \Phi(z_1)\Phi(z_2)\rangle$ as $t\to\infty$.

\section{Conclusions}

We have studied a simple disordered model that, despite the fact that it
is non-integrable before disordering, becomes integrable, by the use of
the replica trick. The theory thus obtained is well defined for all
positive values of the width $\Delta$ of the background field
distribution, and presents no poles as the replica number $n \to 0$.

This example has the interesting feature of being a model with a line of 
defect which is amenable by the methods of integrable field theory, with a 
nontrivial interaction at the defect line.

We used the algebraic method to obtain the reflection and transmission
amplitudes for finite $n$, which, together with the matrix elements of the
defect operator and the form-factors of the free-field theory, were used
to compute correlation functions of the disordered model. Verifying these
results through numerical simulations is an interesting problem.

There are several questions that are worth investigating. We have seen
that a given non-integrable field theory can become integrable as we
disorder it. What about disordering integrable field theories? It would be
interesting to study under what conditions a given integrable model will
still be integrable after disorder, at least in the relevant limit of
$n\to0$.

\section{Acknowledgments}
We would like to thank L. Moriconi for discussions and encouragement, and A. De Martino and G. Mussardo for discussions at an early stage of this work. Financial support from CNPq (PROFIX fellowship), CAPES and FAPERJ is gratefully acknowledged.

\newpage

\section*{Appendix}

In this short appendix we collect some of the formulas used to compute correlation functions within the framework of the replica trick method. More information and references can be found in the review by Bernard \cite{b}.

We can compute quenched correlation functions from the correlation functions in the associated replica model. For example, in order to compute ${\overline{\langle {\cal O}(x) \rangle}}$ we can show that
\be
{\overline{\langle {\cal O}(x) \rangle}}=\lim_{n \to 0}\langle {\cal O}_r(x)\rangle \ .
\ee
Similarly we can show that other quenched correlation functions can be computed from limits of appropriate correlation functions in the replica model
\bea
&&{\overline{\langle {\cal O}(x){\cal O}(y) \rangle}}=\lim_{n \to 0}\langle {\cal O}_r(x){\cal O}(y)\rangle \ , \nonumber \\
&&{\overline{\langle {\cal O}(x) \rangle \langle {\cal O}(y) \rangle}}=\lim_{n \to 0}\langle {\cal O}_r(x){\cal O}_s(y)\rangle; \qquad r \neq s \ .
\eea
These formulas are easily established in the following way. One considers the path-integral definition of the quenched correlation functions, for example
\be
{\overline{\langle {\cal O}(x){\cal O}(y)\rangle}} = \int {\cal D}h \exp\left(-\frac{1}{2\Delta}\int h^2 \right)\frac{\int{\cal D}\phi \exp\left(-S\right){\cal O}(x){\cal O}(y)}{\int{\cal D}\phi \exp\left(-S\right)} \ ,
\ee
with analogous definitions for other quenched correlation functions (connected too). Therefore we can compute the connected two-point correlation function ${\overline{\langle {\cal O}(x){\cal O}(y)\rangle}_c}$ using the analogous definition of the formula above
\bea
&&{\overline{\langle {\cal O}(x){\cal O}(y)\rangle}_c}=\int {\cal D}h\exp\left(-\frac{1}{2\Delta}\int h^2\right)\frac{\delta}{\delta J(x)}\frac{\delta}{\delta J(y)} \ln Z[J] \nonumber \\ 
&&\phantom{{\overline{\langle {\cal O}(x){\cal O}(y)}_c}}=
\lim_{n \to 0}\frac{1}{n}
\frac{\delta}{\delta J(x)}\frac{\delta}{\delta J(y)}\int\prod_{r=1}^n\exp\left(-S_{eff}^{(n)}+\int J\sum_{r=1}^n {\cal O}_r \right) \nonumber \\
&&\phantom{{\overline{\langle {\cal O}(x){\cal O}(y)}_c}}=
\lim_{n \to 0}\frac{1}{n}\langle ({\cal O}_1(x)+\ldots+{\cal O}_n(x))({\cal O}_1(y)+\ldots+{\cal O}_n(y))\rangle\nonumber \\
&&\phantom{{\overline{\langle {\cal O}(x){\cal O}(y)}_c}}=
\lim_{n \to 0} \frac{1}{n}(n\langle {\cal O}_r(x){\cal O}_r(y))\rangle +n(n-1)\langle {\cal O}_r(x){\cal O}_s(y)\rangle_{r \neq s})\nonumber \\
&&\phantom{{\overline{\langle {\cal O}(x){\cal O}(y)}_c}}=
\langle {\cal O}_r(x){\cal O}_r(y)\rangle-\langle {\cal O}_r(x){\cal O}_s(y)\rangle_{r \neq s} \ .
\eea
In the above formulas $S_{eff}^{(n)}$ is the effective action obtained after the integration over the random field with $n$ replicas. All the other identities are established in like manner.

\end{document}